\documentclass[a4paper]{jpconf}
\usepackage{graphicx}
\begin{document}
\title{Freezing of spin dynamics and $\omega/T$ scaling in underdoped cuprates}

\author{Igor Sega$^1$ and Peter Prelov{\v s}ek$^{1,2}$}

\address{$^1$ Jo{\v z}ef Stefan Institute, Ljubljana, Slovenia}
\address{$^2$ Faculty of Mathematics and Physics, University of Ljubljana,
Ljubljana, Slovenia} 

\ead{igor.sega@ijs.si}

\begin{abstract}
The memory function approach to spin dynamics in doped
antiferromagnetic insulator combined with the assumption of
temperature independent static spin correlations and constant
collective mode damping leads to $\omega/T$ scaling in a broad
range. The theory involving a nonuniversal scaling parameter is used
to analyze recent inelastic neutron scattering results for underdoped
cuprates. Adopting modified damping function also the emerging central peak
in low-doped cuprates at low temperatures can be explained within the same
framework.
\end{abstract}

\section{Introduction}
It is by now experimentally well established that magnetic static and
dynamical properties of high-$T_c$ cuprates are quite anomalous. Early 
inelastic neutron scattering (INS) experiments on low-doped La$_{2-x}$Sr$_x$CuO$_4$
(LSCO) \cite{keim,kast} revealed that local, i.e., {\bf q}-integrated
dynamic spin response in the normal state (NS) exhibits anomalous  $\omega/T$ scaling,
not reflected in instantaneous spin-spin correlation length $\xi_T$ which
shows no significant $T$-dependence below room temperature. Subsequently  similar
behaviour has been found in a number of other compounds, i.e., in underdoped
YBaCu$_3$O$_{6+x}$ (YBCO) and in Zn-doped YBCO \cite{kast}. 
More recent INS experiments on heavily underdoped (UD) cuprates,
including Li-doped LSCO \cite{bao}, YBCO \cite{stoc1,stoc2,hink1,hink2}, and
Pr$_{1-x}$LaCe$_x$CuO$_{4-\delta}$ (PLCCO) \cite{wils1}, 
confirm the universal features of  anomalous 
NS spin dynamics so that $\omega/T$
scaling is found in a broad range both in ${\bf q}$-integrated 
susceptibility $\chi''_L(\omega)$ \cite{bao,stoc2,hink2} and in
$\chi''_{\bf q}(\omega)$ at the commensurate AFM ${\bf q}={\bf Q}=(\pi,\pi)$
\cite{bao,stoc2,wils1}. 

Typically, $\chi''_{\bf Q}(\omega)$ is a
Lorentzian with the characteristic relaxation rate scaling as $\Gamma =
\alpha T$, but with a nonuniversal $\alpha$ \cite{bao,stoc2,wils1}.
Similarly, 
$\chi''_L(\omega,T)=\chi''_L(\omega,0)f(\omega/T)$ has been used
\cite{kast,wils1,hink1}, with $f(x)= 2/\pi\arctan [A_1x+A_2x^3]$ and
material dependent $A_{1,2}$. It has been also observed that 
at low $T<T_g$ some intensity is gradually transfered into a central peak
(CP) \cite{bao,stoc1} whereas the inelastic response saturates. This  
{\it freezing} mechanism appears to be entirely dynamical in origin since
$\xi_T$ as well as the integrated intensity are unaffected 
by the crossover.

The present authors introduced a theory of spin dynamics in doped AFM
\cite{prel} which describes the scaling behavior as a dynamical
phenomenon based on two experimental observations: a) $\xi_T$ is (almost) independent 
of $T$, and b) the system is metallic with finite spin
collective-mode damping. Then the  system close to AFM naturally exhibits
$\omega/T$ scaling in a wide energy range, with saturation at low-enough
$T$. 

Our starting point in the analysis of recent INS experiments is the
dynamical spin susceptibilty \cite{prel} 
\vspace{-3pt}
\begin{equation}
\chi_{\bf q}(\omega)=\frac{-\eta_{\bf q}}{\omega^2+\omega
M_{\bf q}(\omega) - \omega^2_{\bf q}}\,,
\label{chiq}
\end{equation}
where the  spin stiffness $\eta_{\bf q}\sim 2J\,(\sim 240$~meV) is only weakly $\bf q$-dependent
($J$ is the exchange coupling), $\omega_{\bf
q}=(\eta_{\bf q}/\chi^0_{\bf q})^{1/2}$ is an effective collective mode
frequency, $\chi^0_{\bf q}\!=\chi_{\bf q}(\omega\!=0)$ is
the static susceptibility and $M_{\bf q}$ is (the complex) memory function
containing information on collective mode damping $\gamma_{\bf
q}=M^{\prime\prime}_{\bf q}(\omega)$.  In the NS of cuprates low-frequency
collective modes at ${\bf q} \sim {\bf Q}$ are generally overdamped so that 
$\gamma_{\bf q}>\omega_{\bf q}$. UD cuprates close to the AFM
phase\break\\
\vspace{-4mm}

\begin{figure}[h]
\vspace{3pt}
\begin{minipage}[t]{20pc}
\vspace{0pt}
\includegraphics[width=20pc]{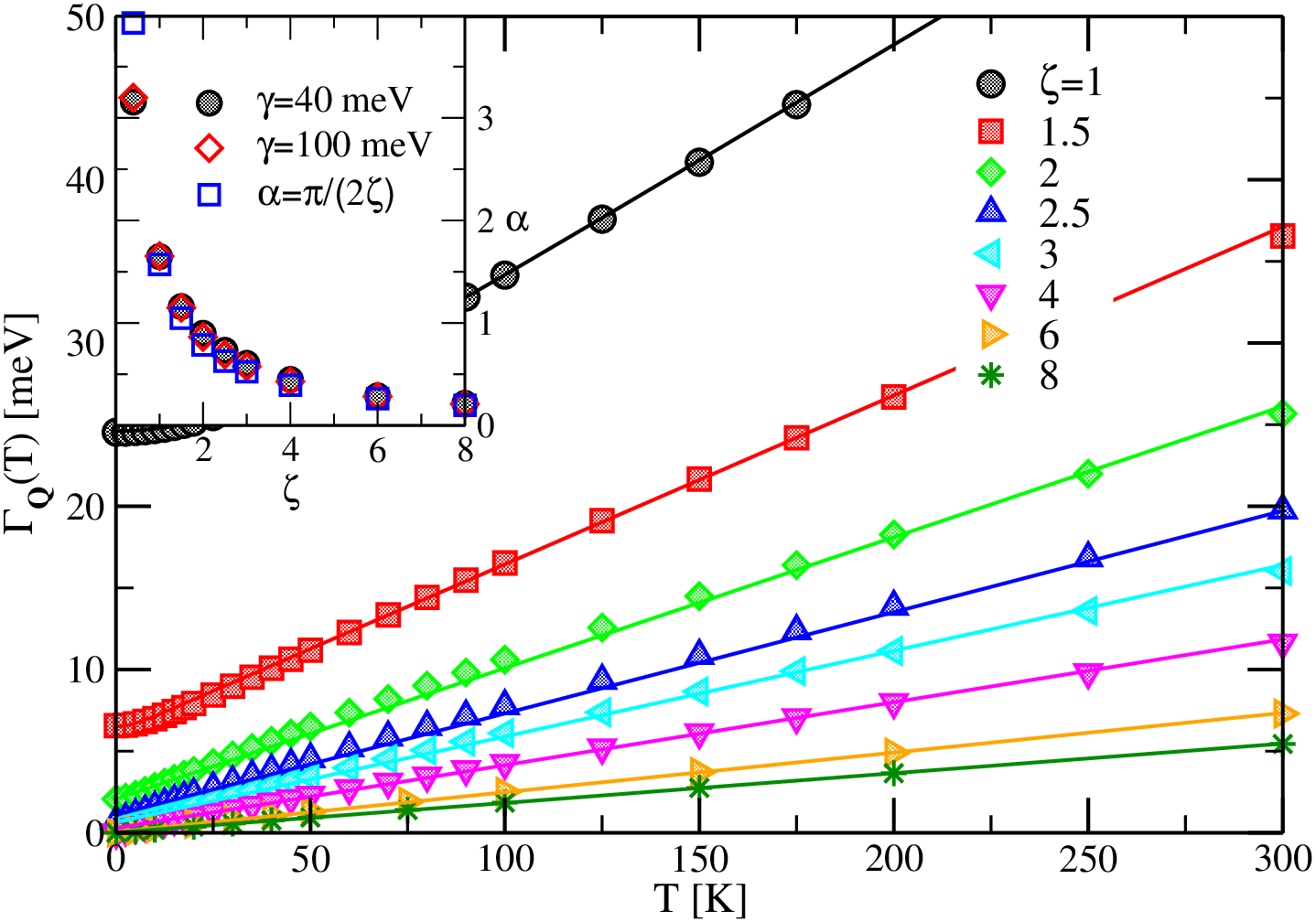}
\caption{\label{fig1}Relaxation rate $\Gamma_{\bf Q}$ vs. $T$ for different
parameters $\zeta$ and $\gamma$=100 meV. Inset: dependence of slope parameter
$\alpha$ on $\zeta$ for two different $\gamma$. For comparison, an estimate
for $\alpha$ is included.}
\end{minipage}
\hspace{2pc}%
\begin{minipage}[t]{15.7pc}
\vspace{-9mm}
have low charge-carrier concentration but pronounced spin
fluctuations whose dynamics is quite generally governed by the sum rule
\begin{equation}
\int_0^\infty \frac{d\omega}{\pi}\chi''_{\bf
q}(\omega)\coth\frac{\omega}{2T}= C_{\bf q}\, ,
\label{eqsum}
\end{equation}
where $C_{\bf q}$ is strongly peaked at ${\bf Q}$ with
a characteristic width $\kappa_T=1/\xi_T$. Moreover, the total sum rule is
for a system with local magnetic moments (spin $1/2$) given by
$(1/N)\sum_{\bf q} C_{\bf q} = (1-c_h)/4$, where $c_h$ is an effective (hole)
doping.

\hspace{1pc}While the formalism so far is very general, we now introduce
approximations specific to UD cuprates \cite{prel}. INS experiments listed above
indicate that within the NS the effective ${\bf q}$ width of
$\chi_{\bf q}''(\omega)$, i.e., dynamical $\kappa(\omega)$, is 
only weakly $T$- and $\omega$-dependent, even on entering the regime with
the\break
\end{minipage}
\vspace{-3pt}
\end{figure}
\vspace{-25pt}
\noindent
CP response \cite{stoc1}. Within further analysis we  assume the commensurate
AFM response at ${\bf Q}$ and the double-Lorentzian form $C_{\bf q}=C/[({\bf
q}-{\bf Q})^2+\kappa_T^2]^2$ although qualitative
results at low $\omega$ do not depend on a particular form of $C_{\bf q}$.

\section{Paramagnetic metal:} 
We also assume that the damping $\gamma_{\bf q}(\omega)$ is
dominated by particle-hole excitations being only weakly ${\bf q}$
and $\omega$ dependent. Hence $\gamma_{\bf q}(\omega) \sim \gamma$, 
with $\gamma$  a phenomenological parameter. 
The assumption of constant $\gamma$ in Eq.~(\ref{chiq}) then leads to
\begin{equation}
\chi''_{\bf q}(\omega) \sim \chi_{\bf q}^0\frac{\omega\Gamma_{\bf q}}
{\omega^2 + \Gamma^2_{\bf q}} , \quad \Gamma_{\bf q}=
\frac{\eta}{\gamma\chi^0_{\bf q}}.
\label{chiim1}
\end{equation}
Note that recent INS data are fully consistent with this form
which has been used to extract $\Gamma_{\bf
Q}(T)$ \cite{stoc2}.

We next exploit the sum rule, Eq.(\ref{eqsum}), to determine $\Gamma_{\bf q}$.
As shown elsewhere \cite{prel,sega2} $\Gamma_{\bf Q}$ is mainly  
determined by the parameter 
\begin{equation}
\zeta=\pi\gamma C_{\bf Q}/(2 \eta), 
\label{zeta}
\end{equation}
subject to $T\ll\gamma$ which is
experimentally relevant. The results are presented in Fig.~1 for a range of
$\zeta=1 - 8$. While $\Gamma_{\bf 
Q}(0)=\Gamma^0_{\bf Q}\sim \gamma {\rm exp}(-2\zeta)$
\cite{prel,sega2}, for $T> \Gamma^0_{\bf Q}$ the variation is nearly linear\break
\noindent
\begin{figure}[t]
\begin{minipage}[t]{240pt}
\includegraphics[width=20pc]{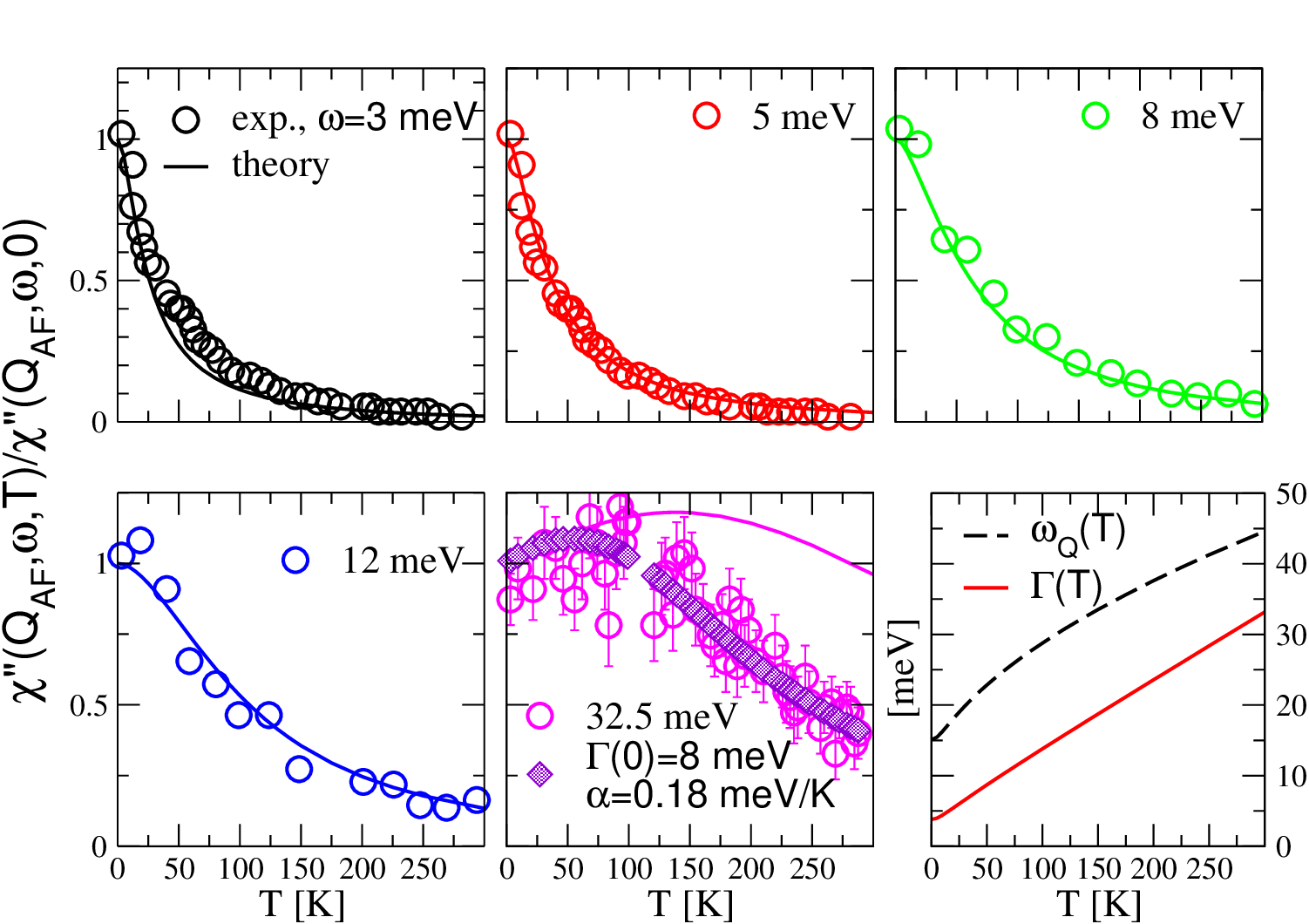}
\caption{\label{fig2}Temperature evolution of $\chi''_{\mathbf Q}(\omega,T)$ for
UD YBCO with $x=0.45$ \cite{hink2} (symbols) compared with
theoretical result, Eq.~(\ref{chiim1}), with $\zeta=1.8$  and $\gamma$=60
meV (lines).}
\end{minipage}\hspace{2pc}%
\begin{minipage}[t]{192pt}
\vspace{-5.85cm}
\caption{\label{fig3}Scaling of theoretical  normalised $\chi''_L(\omega,T)$ based on
$\zeta$ and $\gamma$ as in Fig.~(\ref{fig2}). For comparison, the scaling
function $f(x)$ as frequently used in fits to experiments is also plotted.}
\vspace{15pt}
\includegraphics[width=16pc]{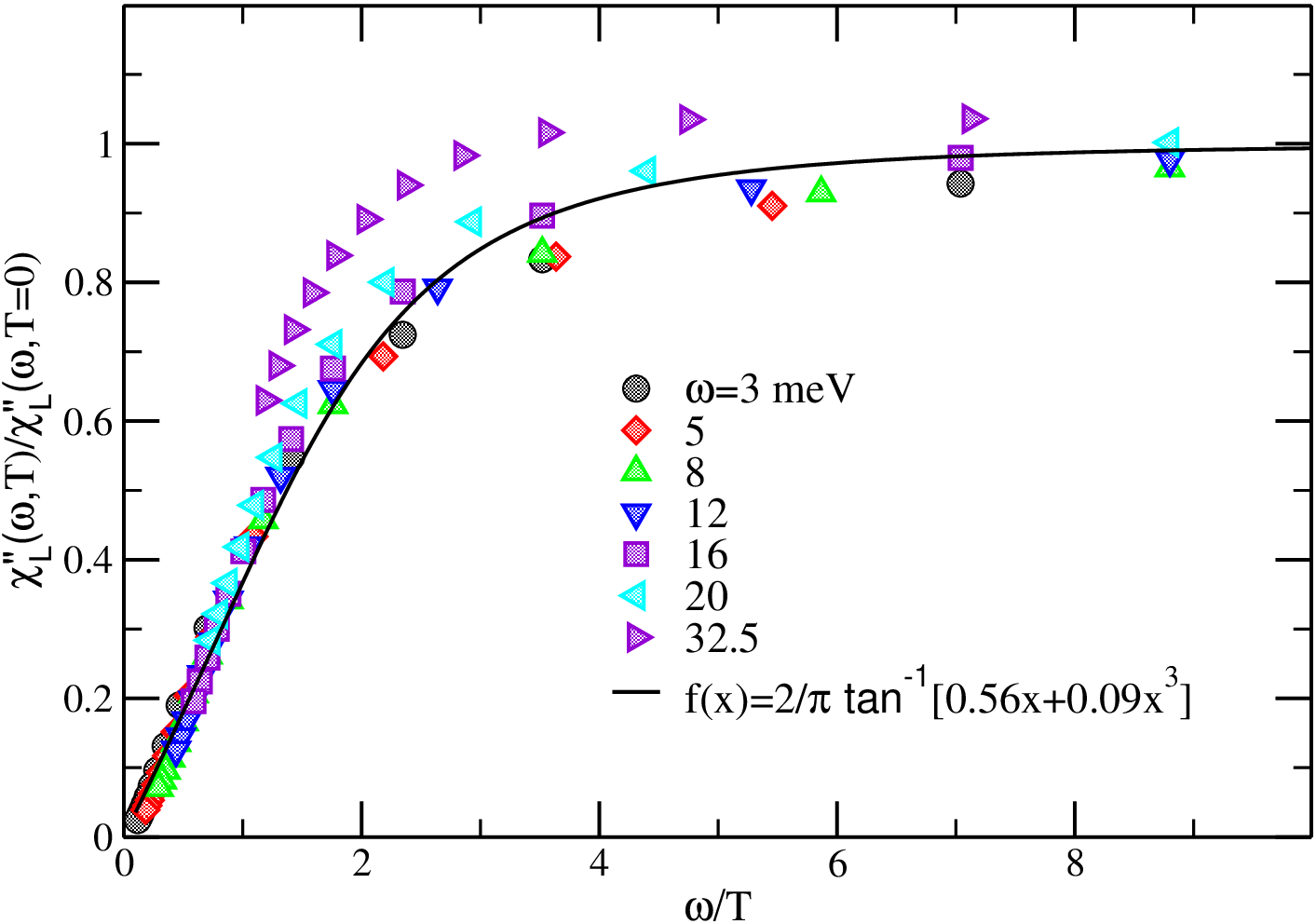}
\end{minipage}
\end{figure}
\vspace{1pc}
\vspace{-36pt}
\begin{figure}[t]
\begin{minipage}[t]{240pt}
\includegraphics[width=20pc]{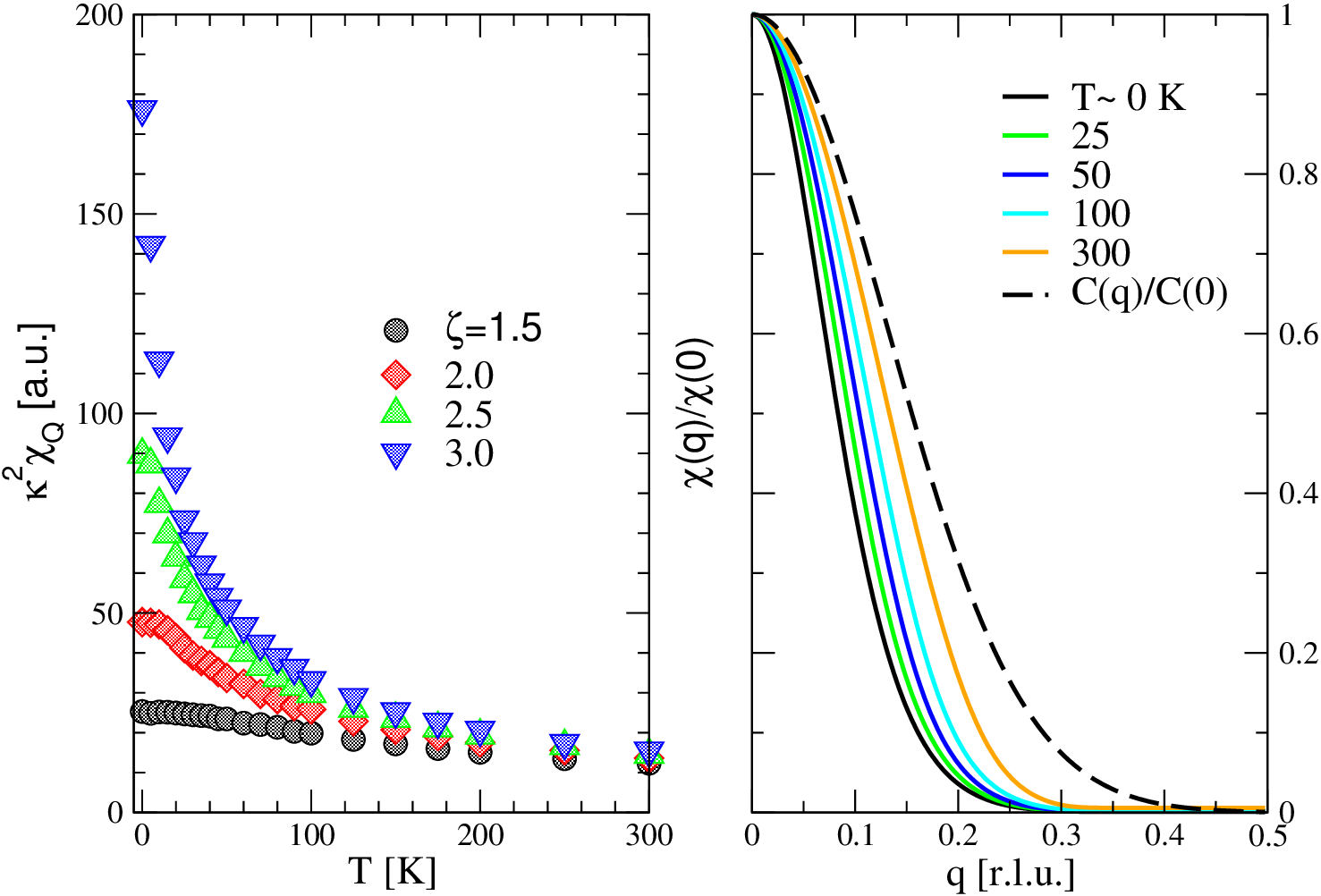}
\caption{\label{fig4}Left pannel: the temperature dependence of
$\kappa^2\chi_{\bf Q}$ for several $\zeta$ as a function of $T$. Right 
pannel: $T$-dependence of $\chi_{\bf q}$ relative to a Lorentzian  $C_{\bf
q}$.}
\end{minipage}\hspace{2pc}%
\begin{minipage}[t]{194pt}
\vspace{-13.5pc}
$\Gamma_{\bf Q} \sim \alpha T$ being a manifestation of the $\omega/T$
scaling. But $\alpha$ is not universal and depends on
$\zeta$ (and weakly on $\gamma$). Then to leading order
\begin{equation}
\Gamma_{\bf Q}(T)\cong\max\{\frac{\pi T}{2\zeta},\gamma\exp(-2\zeta)\},
\label{mfl}
\end{equation}
from which $\alpha=\pi/(2\zeta)$ is identified \cite{sega2}. Note that the
above $\Gamma_{\bf Q}$ leads to 
$\chi''_L(\omega)$ consistent with the   ``marginal Fermi liquid'' model
introduced by Varma et al. \cite{varm}. However, contrary to the usual assumption of
proximity to a quantum critical point where $\xi_T\propto 1/T$,
here $\xi_T\sim$ const.\hfil

\hspace{1pc}Recently a number of INS experiments have been reported where
$\chi''_{\bf Q}(\omega)$ can
be well described by Eq.\ref{chiim1}, including
linear-in-$T$ behaviour of $\Gamma_{\bf Q}$ where,\break
\end{minipage}
\vspace{-3pc}
\end{figure}

\noindent depending on material
$\alpha\sim 0.18 - 0.75$ \cite{bao,stoc2,wils1}. All these $\alpha$ require
rather large $\zeta$, which implies 
very low saturation $\Gamma^0_{\bf Q}$, whereas INS data for
YBCO and LSCO indicate quite substantial $\Gamma^0_{\bf Q}$. However,
saturation of $\Gamma_{\bf Q}$, setting in for $T<T_g$, is 
accompanied by simultaneous appearance of the CP which absorbs the `missing'
sum rule. 

In Fig.~2 INS measurements by Hinkov et al. \cite{hink1,hink2} on UD YBCO 
with $x=0.45$ \cite{hink2} are presented together with
theoretical curves were the only relevant parameter is $\zeta=1.8$, while
in  Fig.~3 theoretical scaling function $\chi''_{\bf
Q}(\omega,T)/\chi''_{\bf Q}(\omega,0)$ for the same $\zeta$ (and $\gamma$)
is plotted. The overall agreement   
between theory and experiment (Fig.~2) is quite satisfactory. The agreement
is less satisfactory  for $\omega=32.5\,$meV and should be attributed to
the breakdown of scaling since $\omega_{\bf Q}>\gamma/2$, as also evident in Fig.~3.
Note that the {\it ad hoc} ansatz for $f(x)$ commonly used (with $A_2=0$) can be easily
obtained assuming a Lorentzian dependence of $\chi_{\bf q}^0$ on $\bf q$. A simple
calculation yields $\arctan(A_1\omega/T)$ with $A_1\sim 1/\alpha$,
provided that  $\kappa\ll1$ but $\kappa^2\chi^0_{\bf Q}\sim$ const (see
Fig.~4). 

\section{CP response:} The advantage of the memory-function
formalism is that the emergence of the CP at $T<T_g$ in the
spin response can be as well treated within the same framework.  One
has to assume that unlike in a paramagnet the mode damping
$\gamma_{\bf q}$ is not constant but may acquire an additional low frequncy
contribution.  In particular we can take
$\tilde M_{\bf q} \sim i \gamma - \delta^2/(\omega+i\lambda)$,
with $T$-dependent $\delta$ and $\lambda$, which leads
to  $\chi''_{\bf q}(\omega)$ of the form used also to analyse experimental
INS data  for YBCO with $x=0.35$ \cite{stoc1,halp}. 

For $\lambda\to0$ but $\delta^2/\lambda\gg \gamma$ the modified $\tilde
M_{\bf q}$ leads to two distinct energy scales and hence to two contributions
to spin dynamics, i.e., the CP part $\chi^c_{\bf q}(\omega)$ and
the regular contribution $\chi^r_{\bf q}(\omega)$,
\begin{equation}
\chi^c_{\bf q}(\omega)\sim \frac{\chi_{\bf
q}^0\Gamma_c}{\Gamma_c-i\omega},\,\, \chi^{r}_{\bf
q}(\omega)\sim \frac{\chi_{\bf q}^{r0}\Gamma_r}{\Gamma_r-i\omega},
\,\,{\bf q}\sim {\bf Q},
\label{cpeq}
\end{equation}
valid for $\omega<\lambda$ and $\lambda<\omega\ll\gamma$,
respectively. Thus, below $T_g$  new scales are set by
$\Gamma_r=\Omega_{\bf q}^2/\gamma$
and $\Gamma_c=(\eta/\delta^2)\lambda/\chi_{\bf q}^0$, with $\chi_{\bf
q}^{r0}=\eta/\Omega_{\bf q}^2$ and $\Omega_{\bf q}^2=\omega_{\bf
q}^2+\delta^2$. If one assumes that $\delta$ saturates at low $T$, as is
manifest by saturation of $\Gamma_r$ \cite{stoc2},
$\Gamma_c\!\propto\!\lambda/\chi_{\bf q}^0$ becomes
the smallest energy scale, resulting in a quasielastic peak of width
$\Gamma_c$. The saturation of $\Gamma_r$, although at present unclear physically,
is responsible for the  transfer of spectral weight, since $C_{\bf
q}^r\sim C_{\bf q}-\pi T/(\gamma\Gamma_r)$ \cite{sega2}, which is 
again consistent with experiment on $x=0.35$ YBCO \cite{stoc2}.

\section{Conclusions} 
The approach presented gives a
consistent explanation of the $\omega/T$ scaling both in $\chi''_{\bf
q}(\omega)$ as well as in $\chi_L''(\omega)$. It is  based on two well
established experimental facts: the overdamped nature of the response and
the saturation of $\kappa_T$ at low $\omega$ and $T$. 
The appearance of the CP for $T<T_g$ is easily incorporated into the
formalism via the (almost) singular $\omega$- and $T$-dependent damping
${\tilde M}_{\bf q}(\omega)$. However, the question as to the origin of CP
remains to be settled.

\end{document}